\documentclass[showpacs,floatfix,twocolumn,prb,superscriptaddress]{revtex4-1}

\usepackage{graphicx,color,upgreek,soul}

\begin{document}

\title{Two-step magnetic ordering into a canted state in ferrimagnetic monoclinic Mn$_3$As$_2$}

\author{Manohar H. Karigerasi}
\author{Bao H. Lam}
\affiliation{Department of Materials Science and Engineering and Materials Research Laboratory, University of Illinois at Urbana-Champaign, Urbana, IL 61801, USA}

\author{Maxim Avdeev}
\affiliation{Australian Centre for Neutron Scattering, Australian Nuclear Science and Technology Organisation, Kirrawee 2232, Australia}

\author{Daniel P. Shoemaker}\email{dpshoema@illinois.edu}
\affiliation{Department of Materials Science and Engineering and Materials Research Laboratory, University of Illinois at Urbana-Champaign, Urbana, IL 61801, USA}


\begin{abstract}
We report the magnetic structure of monoclinic Mn$_3$As$_2$ at 3~K and 250~K using neutron powder diffraction measurements. From magnetometry data, the Curie temperature of Mn$_3$As$_2$ was confirmed to be around 270~K. Calorimetry analysis showed the presence of another transition at 225~K. At 270~K, Mn$_3$As$_2$ undergoes a $k = 0$ ferrimagnetic ordering in the magnetic space group $C2/m$ (\#12.58) with Mn moments pointing along $b$. Below 225~K, there is a canting of Mn moments in the $ac$ plane which produces a multi-$k$  non-collinear magnetic structure in space group $C2/c$ (\#15.85). The components of Mn moments along $b$ follow $k=0$ ordering and the components along $a$ and $c$ have $k = [0 0 \frac{1}{2}]$ propagation vector. The change in the magnetic ground state with temperature provides a deeper insight into the factors that govern magnetic ordering in Mn-As compounds.

\textbf{Keywords} - magnetic structure refinement; neutron diffraction; metallic ferrimagnet; geometric frustration; non-collinear magnetic ordering; magnetocrystalline anisotropy
\end{abstract}



\maketitle 

\section{Introduction} 
The Mn-As phase diagram contains a rich collection of phases with various magnetic structures.\cite{Bacon1955,Yuzuri1960,Carrillo-Cabrera1983,Dietrich1990,Moller1993,Hagedorn1994,Hagedorn1995} Most of the known compounds in this phase-space can be roughly divided into two groups. Compounds in one group are of the form Mn$_{2+n}$As$_{1+n}$ where, starting with stripes of square-planar Mn-As units running along $a$ at $n=0$, every additonal Mn-As involves adding an Mn-As octahedral unit in between the stripes.
In this series, monoclinic Mn$_3$As$_2$ and Mn$_4$As$_3$ correspond to $n=1$ and $2$, respectively. 
It also includes both phases of MnAs where $n=\infty$. The other group consists of tetragonal Mn$_2$As, both the high temperature phases of Mn$_3$As$_2$ and Mn$_5$As$_4$. The structures in this group can be built by constructing slabs from the components of NiAs and Ni$_2$In structure type.\cite{Hagedorn1995} Mn$_3$As and an orthorhombic Fe$_2$P structure type Mn$_2$As are few other compounds that exist in the phase space.\cite{Jeitschko1972,Carrillo-Cabrera1983} 

MnAs orders ferromagnetically (FM) with the Mn moments pointing perpendicular to $c$.\cite{Bacon1955} It changes from a hexagonal NiAs type to an orthorhombic MnP type upon change in temperature, pressure, magnetic field or chemical doping.\cite{Glazkov2003,Ishikawa2006,Sirota1971,Pytlik1985,Schwartz1971} 
The FM ordering of MnAs changes to a spiral or a canted antiferromagnetic (AFM) structure at low temperatures and high pressures.\cite{Bacon1955,Andresen1984,Glazkov2003} 
Mn$_2$As, on the other other hand, has an AFM ordering with N\'eel vector perpendicular to $c$.\cite{Austin1962}
Despite the presence of many compounds in the Mn-As phase diagram, the magnetic structures have been studied only for MnAs and Mn$_2$As.\cite{Bacon1955,Austin1962}
Most known Mn-As compounds provide a metallic lustre upon cleaving.\cite{Hagedorn1995,Hagedorn1994,Dietrich1990,Moller1993}
With increasing interest in metallic antiferromagnets for spintronic applications,\cite{Baltz2018,Siddiqui2020,Jungfleisch2018} the Mn-As phase space provides an ideal collection of compounds to explore magnetism.

\begin{figure}
\centering\includegraphics[width=\columnwidth]{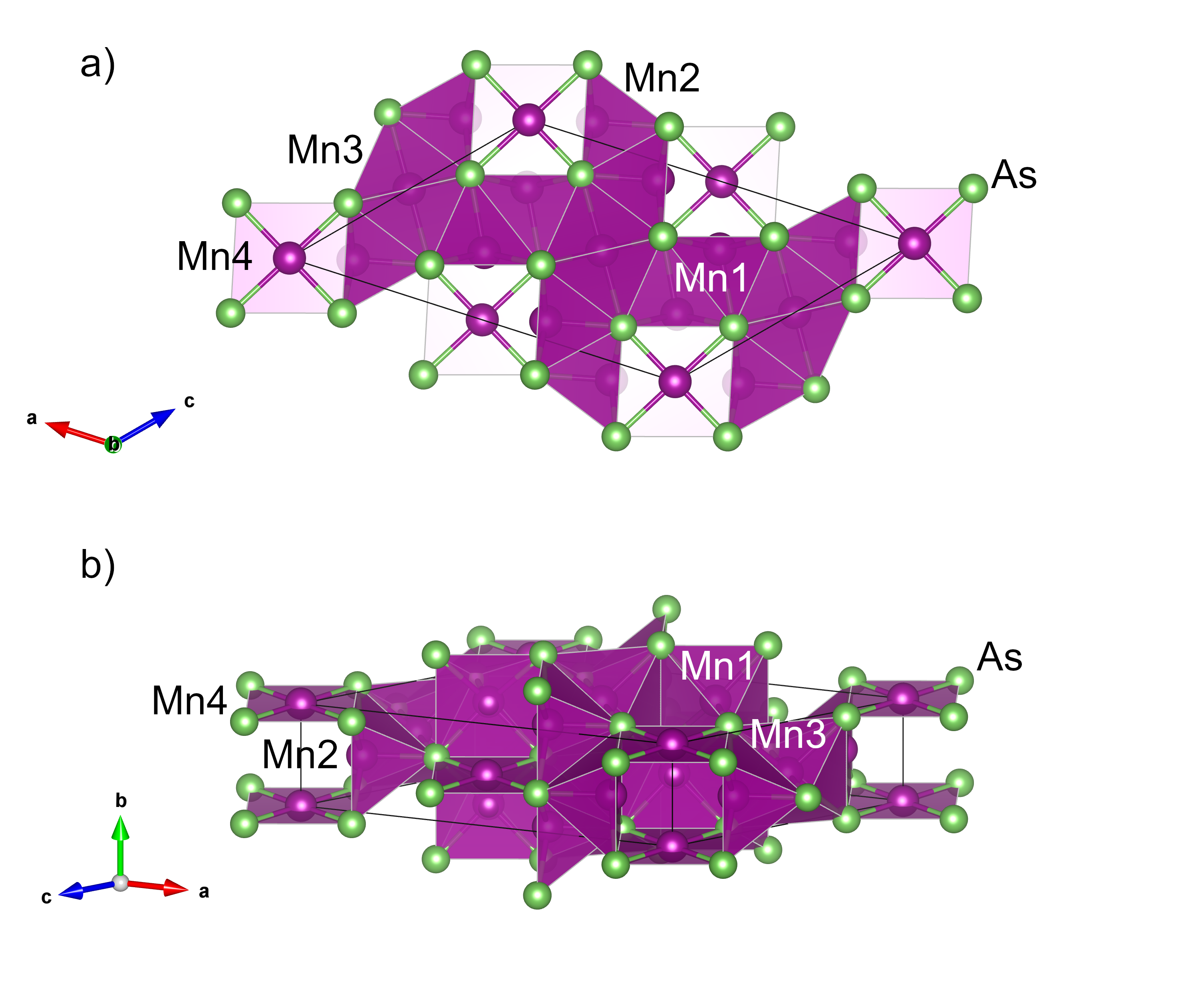} \\
\caption{The chemical structure of Mn$_3$As$_2$ showing the four different Mn atom sites when viewed (a) along $b$ and (b) perpendicular to $b$. Mn1 and Mn2 form square pyramidal units with As, while Mn3 and Mn4 form octahedral and square planar units, respectively.
}
\label{fig:Mn3As2_crystal_structure}
\end{figure}

Mn$_3$As$_2$ is known to exist in three different structure types depending on the stoichiometry and the synthesis procedure.\cite{Dietrich1990,Moller1993,Hagedorn1994} The first variant is in monoclinic space group, which is obtained by quenching after annealing above 1023 K for 9-12 days, and contains a deficiency of Mn atoms. Transport measurements indicate that the compound is metallic.\cite{Dietrich1990} The second variant of Mn$_3$As$_2$ is in orthorhombic space group but the structure can be derived from the previous variant by changing one of the building block in Ni$_2$In structure type.
It is obtained by annealing between 873 K to 1023 K for 9-12 days and is always found to be intergrown with Mn$_5$As$_4$ crystals.\cite{Moller1993} The final variant is the structure that is stable at room temperature when Mn and As are mixed stoichiometrically. 
Single crystal needles of length 0.2~mm can also be obtained with I$_2$ as a transporting agent.\cite{Hagedorn1994}
It crystallizes in a monoclinic space group $C2/m$ with four inequivalent Mn atoms as shown in Figure \ref{fig:Mn3As2_crystal_structure}. Mn atoms form square planar, square pyramidal and octahedral units with As and the structure is very similar to that of tetragonal V$_3$As$_2$.\cite{Hagedorn1994,Hagedorn1995} Magnetometry measurements have indicated that the compound is ferromagnetic below 273 K and the moments saturate at 17.2~gauss per gram or 0.31~$\mu_B$ per Mn atom at low temperature.\cite{Yuzuri1960} 

In this article, we grow room temperature stable monoclinic Mn$_3$As$_2$ using solid state synthesis and carry out magnetometry and differential scanning calorimetry (DSC) measurements to determine the transition temperatures. Using neutron powder diffraction (NPD) measurements, we identify two steps in the magnetic ordering of Mn$_3$As$_2$ and  investigate the crossover from a uniaxial to a canted magnetic  ordering.

\section{Methods}

Bulk polycrystalline Mn$_3$As$_2$ was synthesized by mixing Mn (99.98\% metals basis) and As (99.9999\% metals basis) powders in 3.1:2 ratio using a mortar and pestle inside an Ar filled glovebox. The powders were transferred into a quartz tube, vacuum sealed and heated to 873 K at 2~K/min and held for 2 hours, followed by a ramp at 1~K/min to 1273~K for 1~hour. The sample was then cooled to 1123~K at 1~K/min and held for 1~hour before it was furnace-cooled down to room temperature. The purity of the compound was checked using synchrotron powder x-ray diffraction measurements at the 11-BM beamline of the Advanced Photon Source in Argonne National Laboratory as shown in Figure S1.\cite{supplement} The final product obtained was a solid ingot that was dark gray in color with a metallic luster. Secondary electron images of the crushed Mn$_3$As$_2$ ingot were taken using JEOL JSM-6060LV low-vacuum scanning electron microscope as shown in Figure S2(a) and (b).\cite{supplement}

\begin{figure}
\centering\includegraphics[width=\columnwidth]{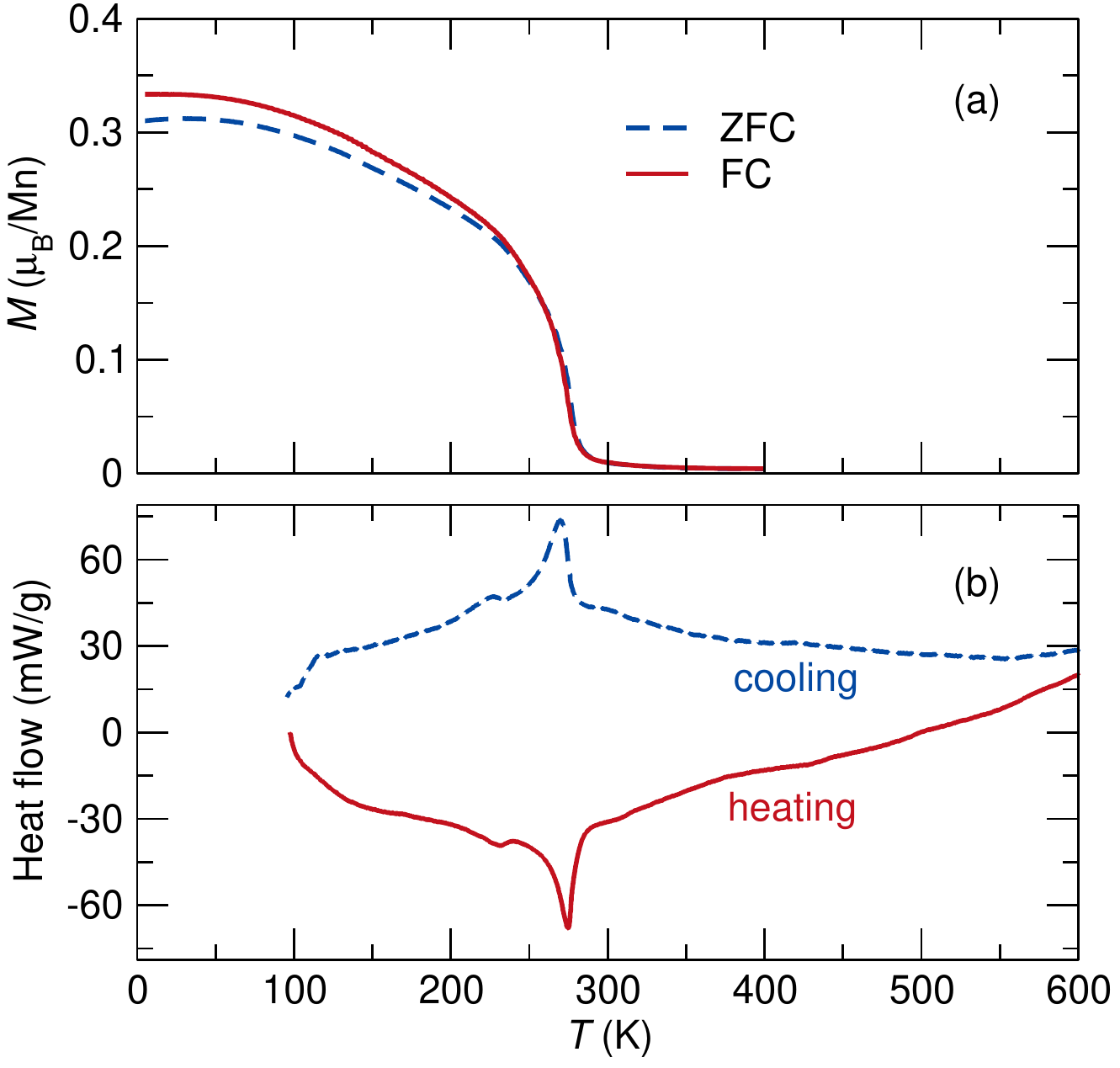} \\
\caption{Field cooling (FC) and zero field cooling (ZFC) of Mn$_3$As$_2$ powders in the presence of 10~kOe field clearly shows a ferromagnetic transition at around 270~K in (a). Heating and cooling curves from the DSC data in (b) show the two transitions at around 270~K and 225~K.
}
\label{fig:DSC_SQUID_measurement}
\end{figure}

\begin{figure}
\centering\includegraphics[width=\columnwidth]{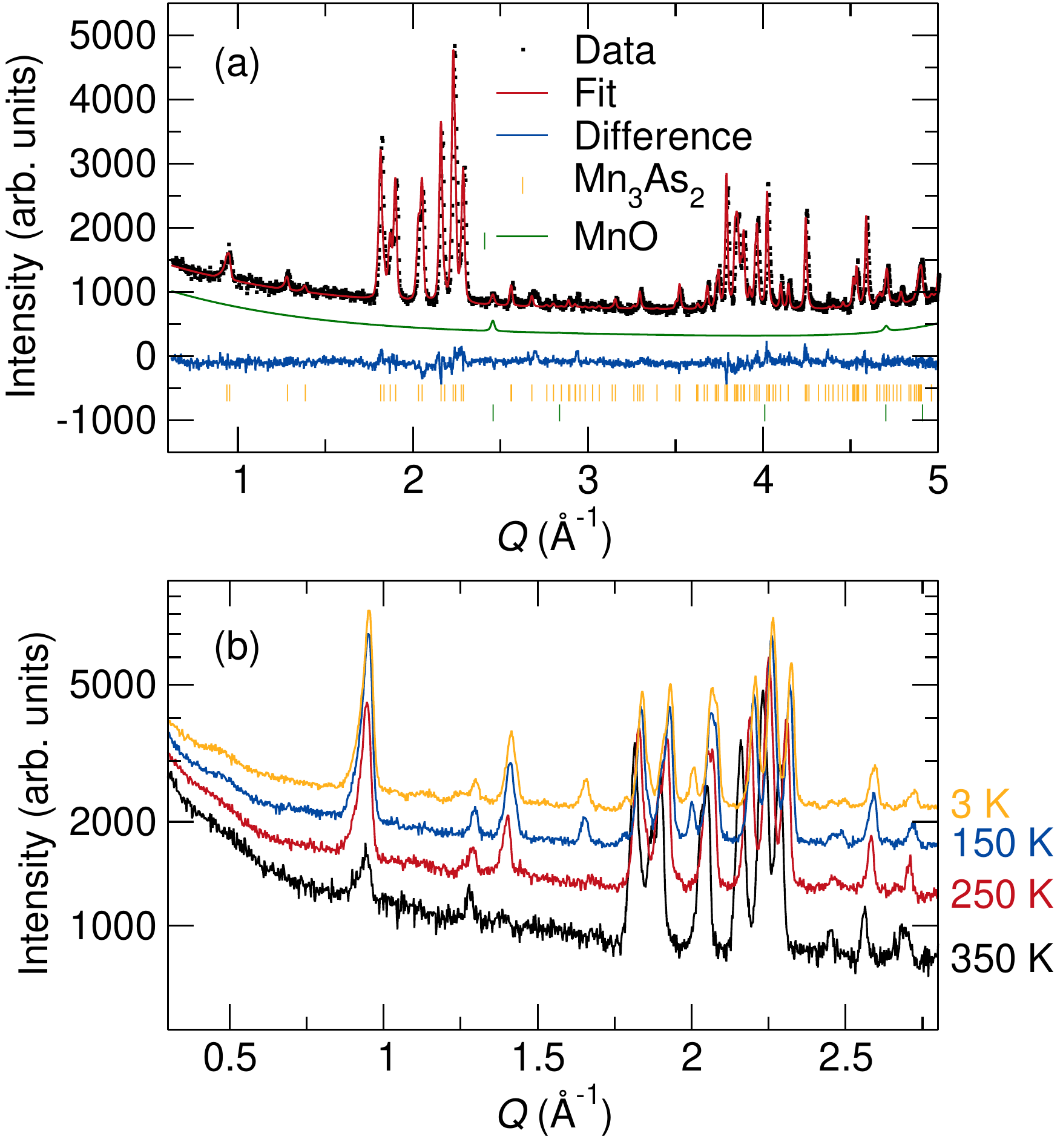} \\
\caption{Rietveld fit to the Mn$_3$As$_2$ NPD data at 350 K is shown in (a). The contribution from the MnO impurity to the fit is also shown. The change in the NPD data due to magnetic transitions upon cooling from 350~K to 3~K is shown in (b). At $T_C = 270$~K, the intensity grows noticeably in the lowest-angle peak, while new peaks appear at the spin-canting transition around 225 K at $Q = 1.65$~\AA$^{-1}$ and 2.0~\AA$^{-1}$.
}
\label{fig:350K_data}
\end{figure}

DSC measurement was carried out on 3.6 mg of powdered sample using Al pans under N$_2$ atmosphere in a TA Instruments DSC 2500. The sample was subjected to a heat-cool-heat cycle between 93~K and 673~K at 10~K/min rate.
Magnetometry was performed on 30.6~mg of powder in a snap-shut sample holder in a Quantum Design MPMS3. The sample was cycled between 400~K and 5~K at 5~K/min in the presence of 10~kOe magnetic field for measuring field cooling (FC) and zero field cooling (ZFC) curves. NPD measurements were carried out on 1.13~g of Mn$_3$As$_2$ powder at the \textsc{ECHIDNA} high resolution powder diffractometer\cite{Avdeev2018} at the Australian Centre for Neutron Scattering. The measurements were done at 3~K, 150~K, 250~K and 350~K. Magnetic structure refinement was carried out using the \textsc{GSAS-II} software\cite{Toby:aj5212} and the \textsc{k-Subgroupsmag} program\cite{Perez-Mato2015} available at the Bilbao Crystallographic Server.


\section{Results and Discussion}



FC and ZFC curves in Figure \ref{fig:DSC_SQUID_measurement}(a) show a clear onset of local magnetic moments near 270~K and the saturation magnetization of 0.33~$\upmu_B$/Mn for field cooling is very close to the reported value of 0.31~$\upmu_B$/Mn.\cite{Yuzuri1960} The Curie temperature was also confirmed by DSC measurements in Figure \ref{fig:DSC_SQUID_measurement}(b). Surprisingly, another transition at around 225~K was observed in the DSC data. This transition is not obvious in the magnetometry data although there seems to be splitting of the FC and ZFC curves at around 225~K. To determine the nature of this transition, whether structural or magnetic, neutron powder diffraction was carried out on these samples at varying temperatures.

\begin{figure}
\centering\includegraphics[width=\columnwidth]{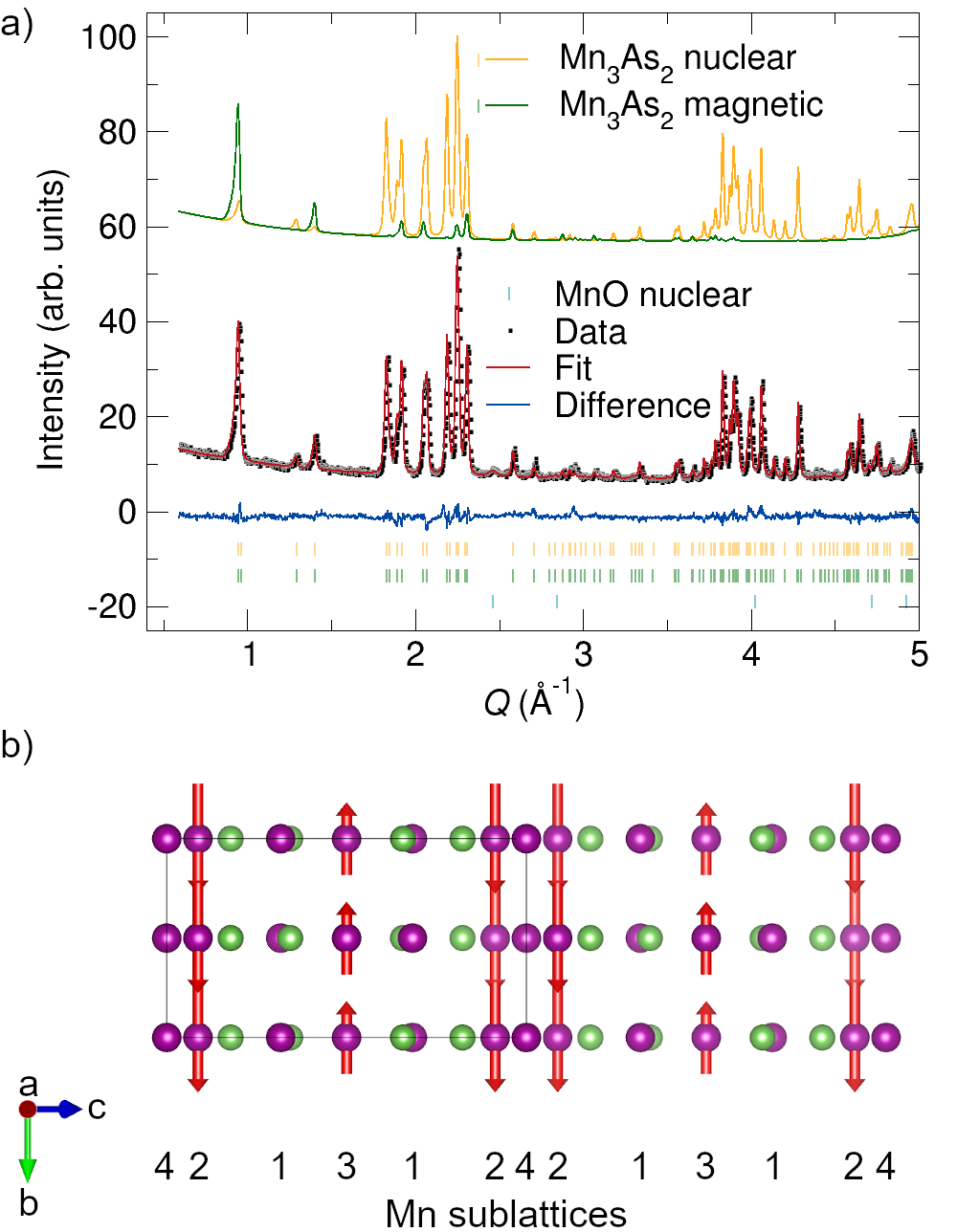} \\
\caption{The Rietveld fit with nuclear and magnetic contributions to the NPD data at 250~K is shown in (a). In (b), the refined magnetic structure is shown. All Mn moments point along $b$ (Mn1 and Mn4 moments are small and along -$b$ and +$b$ directions respectively) and the propagation vector is $k = 0$. 
}
\label{fig:250K_data}
\end{figure}

\begin{figure}
\centering\includegraphics[width=\columnwidth]{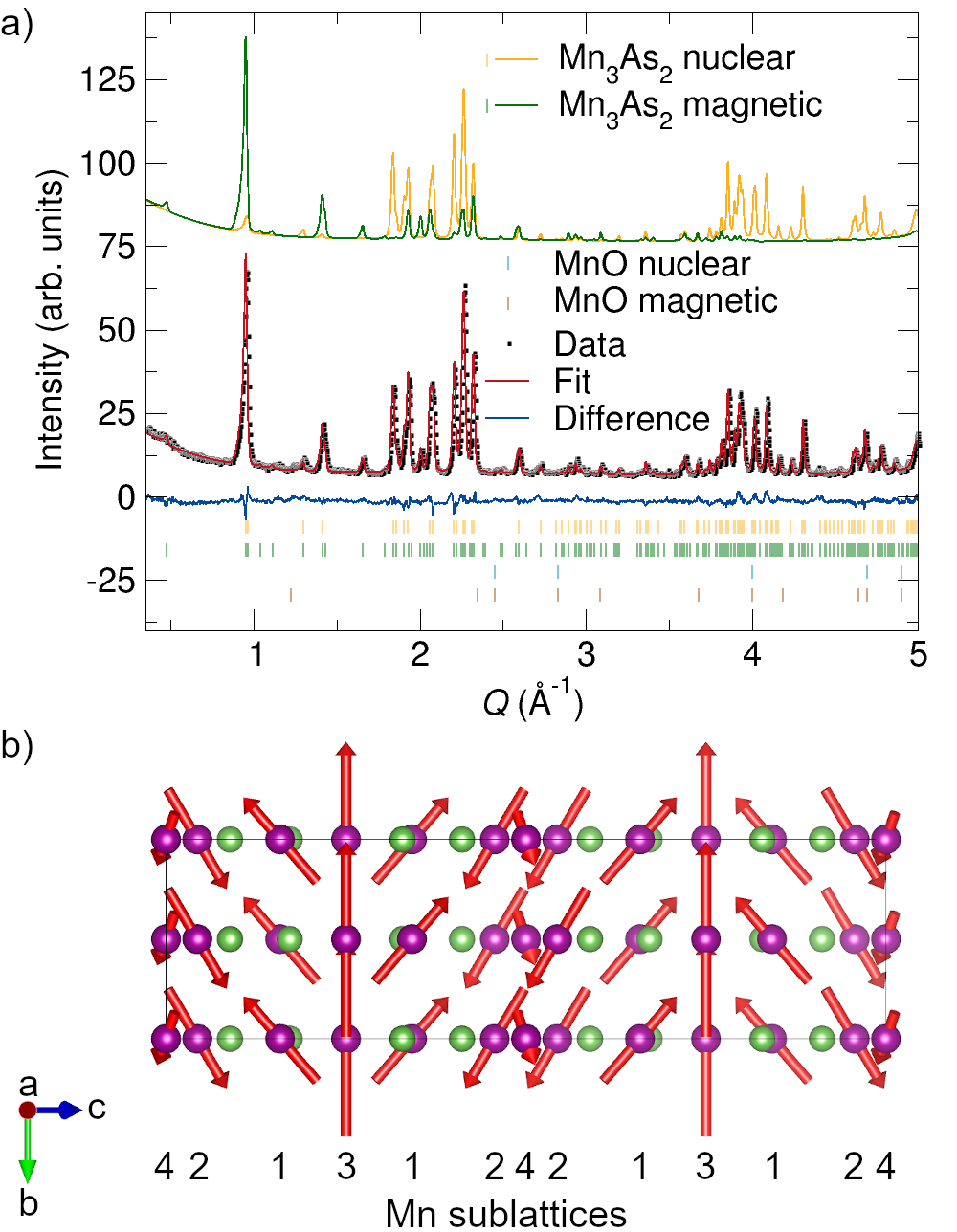} \\
\caption{Rietveld fit (a) to the NPD data at 3~K for the magnetic structure shown in (b). There is a canting of spins in the $a-c$ plane which results in multiple magnetic ordering vectors $k = 0$ and $k = [00\frac{1}{2}]$. The nuclear and the magnetic contribution to the fit is also shown in (a). 
}
\label{fig:3K_data}
\end{figure}


The Rietveld fit to the NPD data at 350 K in Figure \ref{fig:350K_data}(a) confirms the paramagnetic nature of Mn$_3$As$_2$ above 270~K. About 0.6 wt\% of MnO was present as impurity and its contribution to the NPD data is shown in Figure \ref{fig:350K_data}(a). Subsequent Rietveld fits to the NPD data at different temperatures also account for the nuclear contribution from the MnO impurity phase. Cooling from 350~K to 3~K introduces additional peaks that are magnetic in nature as seen in Figure \ref{fig:350K_data}(b).
At 250 K, the intensities of the peaks near $Q = 0.95$~\AA$^{-1}$ and 1.4~\AA$^{-1}$ increase considerably. Since both peaks are not structurally forbidden, the magnetic unit cell remains the same as the chemical unit cell.
At 150~K, we can see new magnetic peaks near $Q = 1.65$~\AA$^{-1}$ and 2.0~\AA$^{-1}$. Fits to these patterns confirm the kink observed in the DSC data at 225~K to be a magnetic transition. The NPD patterns remain consistent upon further cooling, so we are confident the magnetic structure does not change between 150~K and 3~K. The T$_N$ of MnO is 120~K but the intensity from the magnetic peaks of MnO are too weak to be observed here. MnO magnetic peaks do not overlap with any of the Mn$_3$As$_2$ magnetic peaks and we provide markers corresponding to the MnO magnetic peaks at 3~K NPD data.


The magnetic ordering vector of Mn$_3$As$_2$ at 250~K is $k = 0$. The indices of the two magnetic peaks correspond to $(001)$ and $(20\overline{2})$ respectively. Since the magnetic peaks are of the form $(h0l)$, it is likely that the Mn moments would prefer to orient along $b$. In the $C2/m$ space group with propagation vector $k=0$, there are four possible $k$-maximal subgroups that are consistent with this propagation vector. The four subgroups correspond to different combinations of the addition of the time reversal operator to the 2-fold axis and the mirror plane. 
Of the four models, two models restrict the Mn moment orientation to the $b$ axis and the other two restrict Mn moments to lie in the $ac$ plane. 
One model from each pair results in an AFM structure and provides a poor fit to the NPD data. 
The best fit (R$_{wp}$ = 5.087\%) is unambiguously obtained for the model with $C2/m$ space group symmetry where all Mn moments point along the $b$ axis as shown in Table S1.
The refined Mn moments for this ferrimagnetic structure are provided in Table \ref{tab:Mn_moments}. The net moment is 0.43(5)~$\upmu_B$/Mn which is close to the saturation moment of 0.33~$\upmu_B$/Mn from magnetometry. The Rietveld fit of this model to the NPD data and the magnetic structure are shown in Figure \ref{fig:250K_data}(a) and (b), respectively.


\begin{table*}
\caption{\label{tab:Mn_moments} 
The magnetic space groups, propagation vectors (k-vectors), magnetic irreducible representations (mag IRs) and the Mn moments in $\upmu_B$ for the magnetic structures at two different temperatures. 
}
\centering
\begin{tabular}{p{1.5cm}p{2cm}p{2.2cm}p{2.6cm}p{2.15cm}p{2.15cm}p{2.15cm}p{2.15cm}}
\hline\hline
\bf{T (K)} & \bf{MSG} & \bf{k-vectors} & \bf{mag IRs} & \bf{Mn1 (x,y,z)} & \bf{Mn2 (x,y,z)} & \bf{Mn3 (x,y,z)} & \bf{Mn4 (x,y,z)}\\
\hline\hline
250 & $C2/m$ & 0 & mGM$_1^+$ & 0.00 & 0.00 & 0.00 & 0.00\\
& ($\#$12.58) & & & -0.52(9) & 2.55(9) & -1.69(14) & 0.59(6)\\
& & & & 0.00 & 0.00 & 0.00 & 0.00\\
\hline
3 & $C2/c$ & 0, $[00\frac{1}{2}]$ & mGM$_1^+$ + mA$_2^+$ & 1.24(15) & 0.87(15) & 0.00 & -0.91(16)\\
& ($\#$15.85) & & & -1.94(8) & 2.28(7) & -4.48(13) & 1.21(6)\\
& & & & 2.27(11) & 1.82(10) & 0.00 & -0.56(11)\\
\hline\hline
\end{tabular}
~\\
\end{table*}

All magnetic peak locations in the NPD data at 3~K and 150~K can be indexed using a propagation vector of $k = [00\frac{1}{2}]$. However, none of the magnetic structures from the subgroups consistent with this propagation vector provide a good fit to data and all magnetic structures obtained are AFM, inconsistent with magnetometry. Refining the 250~K model to the low-temperature NPD data provides a good fit to the two previously-existing magnetic peaks but none of the additional peaks can be fit using this model. For these reasons, it is clear that below 225~K, Mn$_3$As$_2$ contains two propagation vectors, $k = 0$ and $k = [00\frac{1}{2}]$.

With $C2/m$ as the parent space group and using both propagation vectors, there are 16 possible $k$-maximal subgroups. 
Since the magnetic irreducible representations (irreps) of the 4 $k$-maximal subgroups in each propagation vector are one-dimensional, there is a one to one correspondence between the irreps and the space groups. The 16 $k$-maximal subgroups are obtained by mixing the 4 irreps from one propagation vector with the 4 irreps from the other propagation vector.
Expecting that the low-temperature magnetic ordering is similar to the one at 250~K, we can choose the 4 subgroups that contain the same irrep as the 250~K structure. 
This leaves us with 2 magnetic structures each in $C2/c$ and $C2/m$ space groups. The $C2/m$  magnetic structures contain all Mn moments pointing along $b$, but none of the fits provide required intensity at the $Q = 2.0$~\AA$^{-1}$ magnetic peak. As shown in Table S2, out of the two magnetic structures with $C2/c$ space group, the best fit (R$_{wp}$ = 5.852\%) was obtained for the structure where the Mn3 moment was constrained by symmetry to be along $b$ and all other moments were allowed to tilt away from $b$. Moving to lower symmetry does not justify the additional 7 or 8 variables in the refinement. The magnitudes of the refined Mn moments are given in the Table \ref{tab:Mn_moments}. The magnetic structure along with the refined fit to the NPD data at 3~K is given in Figure \ref{fig:3K_data}. The mcif files for both the magnetic structures are attached as Supplementary Materials.\cite{supplement}

\begin{figure}
\centering\includegraphics[width=\columnwidth]{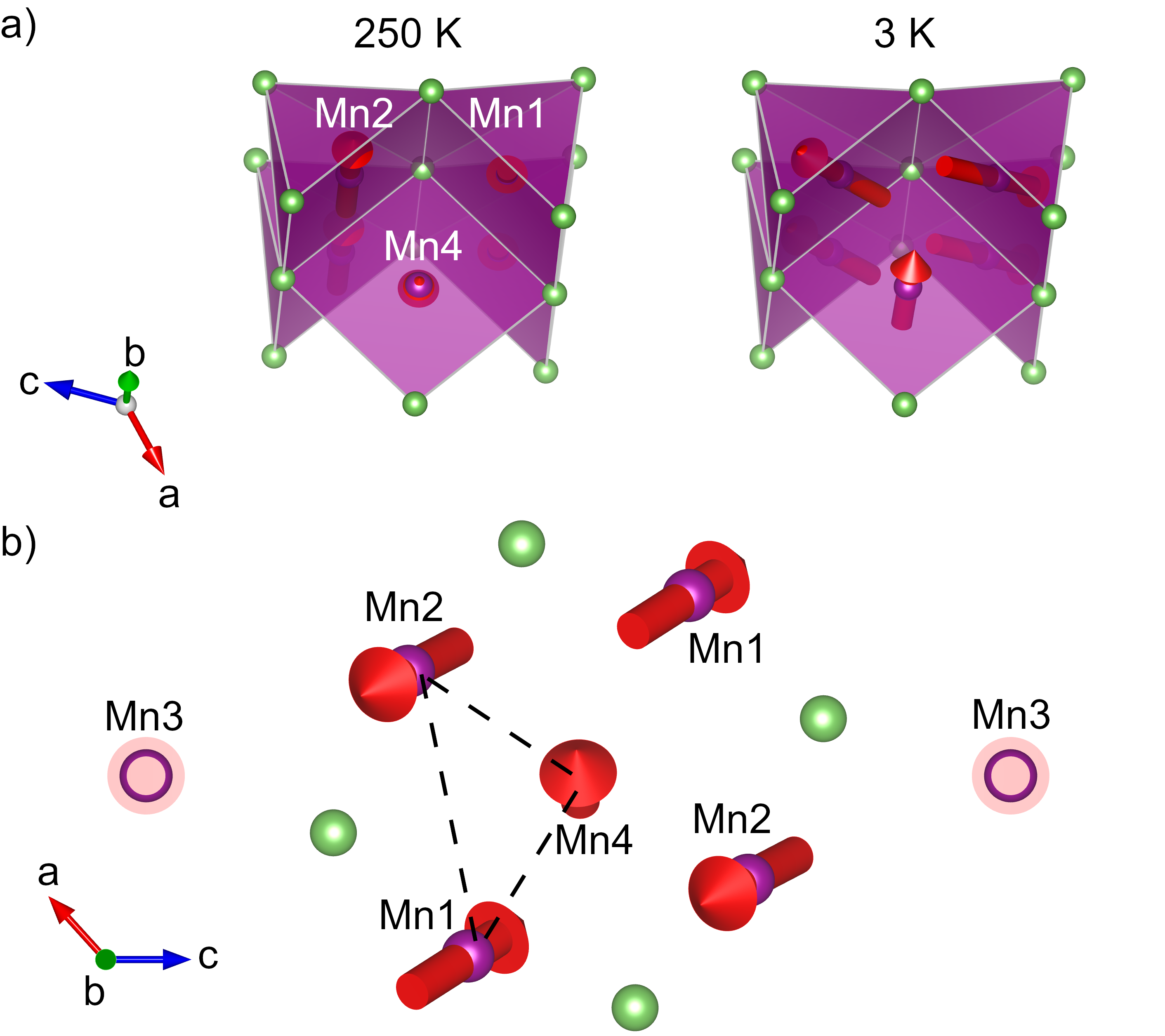} \\
\caption{At 250~K, large moments are present within the basal plane of square pyramidal units of Mn1 and Mn2 atoms (left, (a)). Upon decrease in temperature, there is ordering of Mn4 moments (right, (a)) which induces canting in Mn1 and Mn2 moments as well. (b) shows the geometric frustration due to antiferromagnetic interactions between Mn1, Mn2 and Mn4 moments.
}
\label{fig:spin_canting}
\end{figure}


There are 80 entries having two propagation vectors in the MAGNDATA database,\cite{Gallego2016_1,Gallego2016_2} which is about 7\% of all structures reported in the database. Hence, although less common, it is not rare to find compounds with multi-k structures. However, more than half of the compounds have either a collinear spin arrangement or contain a secondary $k = 0$ ordering to account for the presence of homogeneous magnetic moments.\cite{Klepov2019,Mekata1978} Of the remaining compounds, a common theme is to have multiple propagation vectors (k-vectors) act on different sublattices at different temperatures\cite{Zhang2019,Yi2015,Jin2013} or contain 2 k-vectors belonging to the k-star of the active irrep.\cite{Skanthakumar1991} There are less than ten compounds in the database where multiple irreps from different k-vectors act on the same magnetic sites at different temperatures like in our case.

A quick look at the magnetic transitions in this compound might lead one to compare with the ordering in triangular antiferromagnets with a small uniaxial anisotropy such as CsNiCl$_3$ and CsNiBr$_3$.\cite{Collins1997} Upon cooling from the paramagnetic state in triangular antiferromagnets, the spins are first ordered collinear along the easy axis and then undergo another transition to form a non-collinear frustrated triangular arrangement with one of the three spins pointing along the easy axis. However, the frustration in Mn$_3$As$_2$ is along the $a-c$ plane as shown in Figure \ref{fig:spin_canting}(b) and not $a-b$ or $b-c$ plane as would be expected if it were a triangular antiferromagnet. Hence, the anisotropy in Mn$_3$As$_2$ is not simply uniaxial below 225~K. There is some driving force that favors frustration in the $a-c$ plane which, according to our hypothesis, arises from the anisotropy of Mn4 moments.
In metallic antiferromagnets such as Mn$_2$As, tetragonal and orthorhombic CuMnAs,\cite{Austin1962,MacA2012,Wadley2013} all Mn moments in square pyramidal units with As are oriented within the basal plane rather than along the 4-fold symmetric axis.
Hexagonal Cu$_{0.82}$Mn$_{1.18}$As is a frustrated system and hence, the Mn moments slightly deviate from this arrangement.\cite{Karigerasi2019}
There are no magnetic structures reported in the MAGNDATA database\cite{Gallego2016_1,Gallego2016_2} where Mn forms square planar units with As, as in Mn4 atoms in Mn$_3$As$_2$. The canting of spins in Mn$_3$As$_2$ can be explained if we assume that the Mn spins, when bonded with As in these square planar units, prefer to orient in-plane.

In Mn$_3$As$_2$ at high temperatures, the molecular fields from other Mn moments induces a net moment in square-planar Mn4 along $b$, as shown in Figure \ref{fig:spin_canting}(a). The value is small (0.59~$\upmu_B$) at 250~K from Table \ref{tab:Mn_moments}. Such behavior has also been observed in other arsenides such as Cr$_2$As where the Cr2 sublattice orders first at 393~K and induces a weak moment in the Cr1 atoms. The Cr1 moments order at a much lower temperature at around 175~K.\cite{Ishimoto1995} 
At 225~K, the magnetocrystalline anisotropy of Mn4 moments becomes significant compared to the thermal energy and the moments acquire components along $a$ and $c$. Through exchange interactions with Mn1 and Mn2 moments, there is a canting of the Mn1 and Mn2 moments as well away from $b$, as shown in Figure \ref{fig:spin_canting}(a). The non-collinear arrangement of Mn spins is further enhanced through geometric frustration in the $a-c$ plane due to competing AFM interactions between Mn1, Mn2 and Mn4 moments sitting on a distorted equilateral triangle as shown in Figure \ref{fig:spin_canting}(b). There are not enough data points in the linear paramagnetic regime of the inverse susceptibility curve to provide a Curie-Weiss fit as shown in Figure S3.\cite{supplement}
The coordination environments of Mn atoms in Mn$_3$As$_2$ have point symmetry $m$ and $2/m$, not the highest allowed by their immediate coordination environments (which are distorted), but the single-site anisotropies still seem to broadly obey the trend of basal-plane preference in square pyramids and square planes.  This consistency provides opportunity to design magnetic structures by choosing specific magnetic motifs, even in low-symmetry compounds. 

The symmetry-breaking spin canting in Mn$_3$As$_2$ may at first glance seem surprising, given the nominal Mn$^{2+}$ and $3d^5$ electron configuration, but magnetocrystalline anisotropy in Mn-containing arsenides is  quite complex. Even within a set of compounds with common cation oxidation state and anion character and coordination, the spin-orbit coupling of excited and occupied states plays a major role, and typically requires computational investigation.\cite{Duboc2016}
Among compounds with two propagation vectors, inelastic neutron scattering of SrHo$_2$O$_4$ powders was used to reveal that single-ion anisotropies of the Ho sites could explain the zig-zag chain ordering in the compound.\cite{Fennell2014} TbOOH is another 2-k non-collinear compound that also has a $k = 0$ ordering. From dipolar energy calculations, the non-collinear arrangement was attributed to the crystal field anisotropy of the Tb$^{3+}$ ions which results in anisotropic exchange interactions between Tb ions.\cite{Christensen1974}
The specific energy scales that are relevant in Mn$_3$As$_2$ require further computational work and a broader set of materials to investigate. 

\section{Conclusion}

The magnetic structure of monoclinic Mn$_3$As$_2$ was identified using neutron powder diffraction experiments. From SQUID magnetometry measurements, it was identified that the material is a weak ferromagnet below 270~K. DSC data indicated another transition at around 225~K. From NPD data at 250~K, it was found that Mn$_3$As$_2$ is a ferrimagnet with all Mn moments ordering along $b$. Between 225~K to 270~K, the compound has a $k = 0$ magnetic ordering. Below 225~K, there is a canting of spins in the $a-c$ plane and it has a multi-k ordering structure with an additional $k = [00\frac{1}{2}]$ propagation vector. Here, the component of Mn moments along $b$ follow $k = 0$ ordering and the moments are uncompensated. The component of Mn moments along $a$ and $c$ follow $k = [00\frac{1}{2}]$ ordering. This behavior can be explained by considering that Mn moments align in the plane of the square planar or square pyramidal Mn-As units. Mn4 atoms are bonded to As in square-planar units within the $a-c$ plane. The lower temperature transition simply corresponds to the ordering temperature of the Mn4 sublattice.  Below 225~K, Mn4 moments cause spin canting in Mn1 and Mn2 moments through exchange interactions. Geometric frustration between Mn1, Mn2 and Mn4 moments cause significant deviation from the collinear arrangement of spins.

\section{Acknowledgments}
This work was undertaken as part of the Illinois Materials Research Science and Engineering Center, supported by the National Science Foundation MRSEC program under NSF Award No. DMR-1720633. 
The characterization was carried out in part in the Materials Research Laboratory Central Research Facilities, University of Illinois.
Use of the Advanced Photon Source at Argonne National Laboratory was supported by the U.~S. Department of Energy, Office of Science, Office of Basic Energy Sciences, under Contract No. DE-AC02-06CH11357.

\bibliography{mn3as2_mono}

\end{document}



\begin{center}
\Large 
\textbf{Two-step magnetic ordering into a canted state in ferrimagnetic monoclinic Mn$_3$As$_2$}\\
\vspace{1em}
Supplementary Material\\
\vspace{1em}
\normalsize
Manohar H. Karigerasi, Bao H. Lam, Maxim Avdeev, Daniel P. Shoemaker
\end{center}

\vspace{2em}

\begin{figure}[h]
\centering\includegraphics[width=0.7\columnwidth]{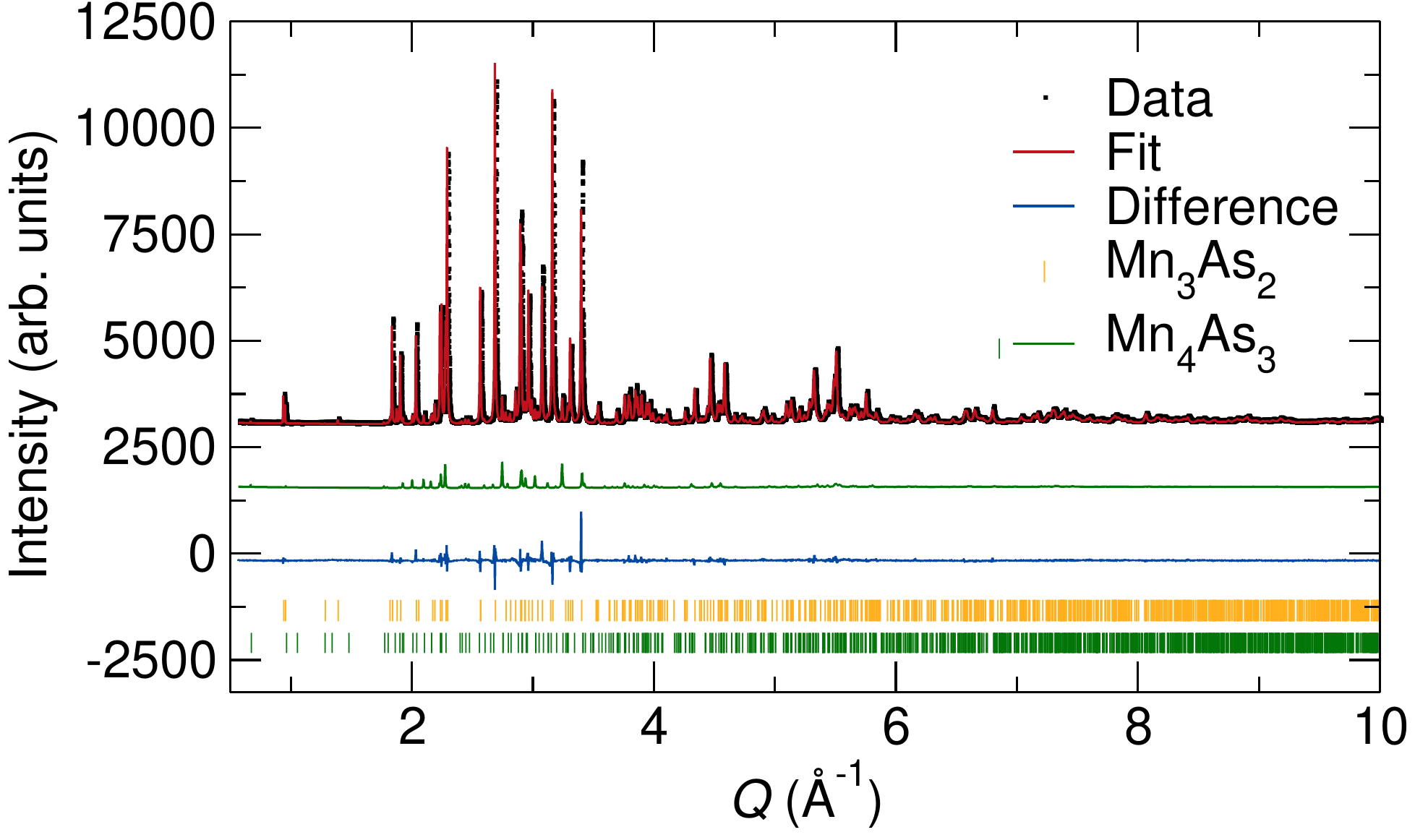} \\
\caption{\label{fig:11BM_data}
Rietveld fit to the synchrotron powder x-ray diffraction data of Mn$_3$As$_2$ showed 7.4 wt.\% Mn$_4$As$_3$ impurity. The contribution of the Mn$_4$A$_3$ impurity phase to the diffraction data is also shown in the figure. This impurity was, however, not seen in the NPD data.
} 
\end{figure}

\begin{figure}
\centering\includegraphics[width=0.7\columnwidth]{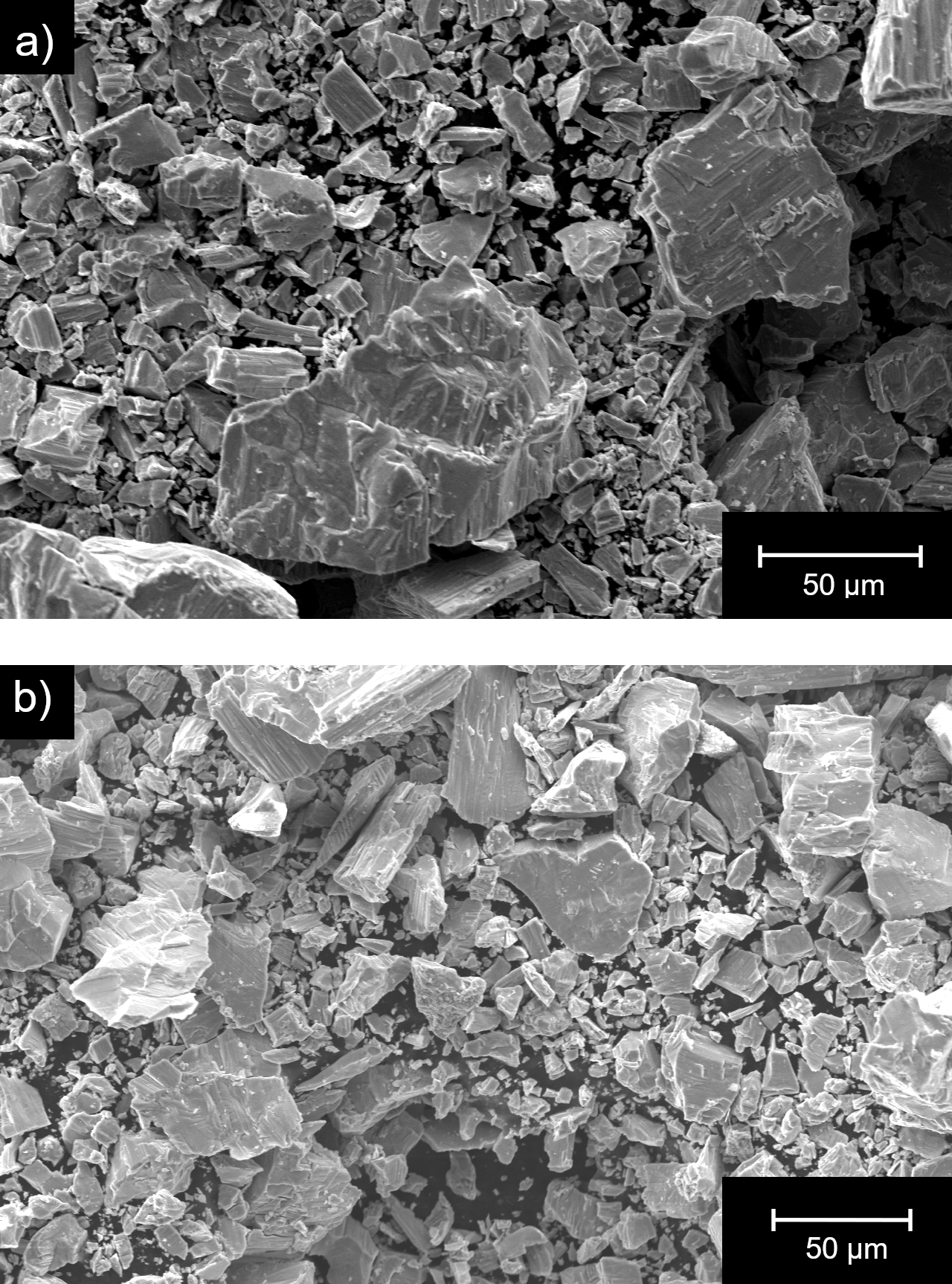} \\
\caption{\label{fig:SEM_image}
Scanning electron microscopy image of Mn$_3$As$_2$ crystals crushed from an ingot is shown in (a) and (b). Clear facets in the crystals indicate melting of the elemental powders during synthesis. 
} 
\end{figure}

\FloatBarrier

\begin{table}
\caption{\label{tab:refinement_250_K} 
The goodness of fit measured in terms of data residual (R$_{wp}$) and the unweighted phase residual of the magnetic phase (RF$^2$) for the four k-maximal subgroups at 250~K. 
}
\centering
\begin{tabular}{p{3cm}p{2.5cm}p{2.5cm}p{2.5cm}}
\hline\hline
\bf{Space group} & \bf{No.} & \bf{R$_{wp}$ (\%)} & \bf{RF$^2$ (\%)}\\
\hline\hline
$C2'/m'$ & $\#$12.62 & 5.954 &  19.579\\
$C2/m'$ & $\#$12.61 & 6.880 &  27.604\\
$C2'/m$ & $\#$12.60 & 6.186 &  21.937\\
\bf{C2/m} & \bf{\#12.58} & \bf{5.087} &  \bf{12.018}\\
\hline\hline
\end{tabular}
~\\
\end{table}

\begin{table}
\caption{\label{tab:refinement_3_K} 
The goodness of fit measured in terms of data residual (R$_{wp}$) and the unweighted phase residual of the magnetic phase (RF$^2$) for the four k-maximal subgroups containing the magnetic irrep mGM$_1^+$ at 3~K.
}
\centering
\begin{tabular}{p{3cm}p{2cm}p{3cm}p{2.5cm}p{2.5cm}}
\hline\hline
\bf{Space group} & \bf{No.} & \bf{Translation vector} & \bf{R$_{wp}$ (\%)} & \bf{RF$^2$ (\%)}\\
\hline\hline
\bf{C2/c} & \bf{\#15.85} & \bf{[0,0,0]} & \bf{5.852} &  \bf{6.773}\\
$C2/c$ & $\#$15.85 & $[0,0,\frac{1}{2}]$ & 6.112 &  8.079\\
$C2/m$ & $\#$12.58 & $[0,0,0]$ & 6.829 &  9.831\\
$C2/m$ & $\#$12.58 & $[0,0,\frac{1}{2}]$ & 6.719 &  8.486\\
\hline\hline
\end{tabular}
~\\
\end{table}

\begin{figure}
\centering\includegraphics[width=0.6\columnwidth]{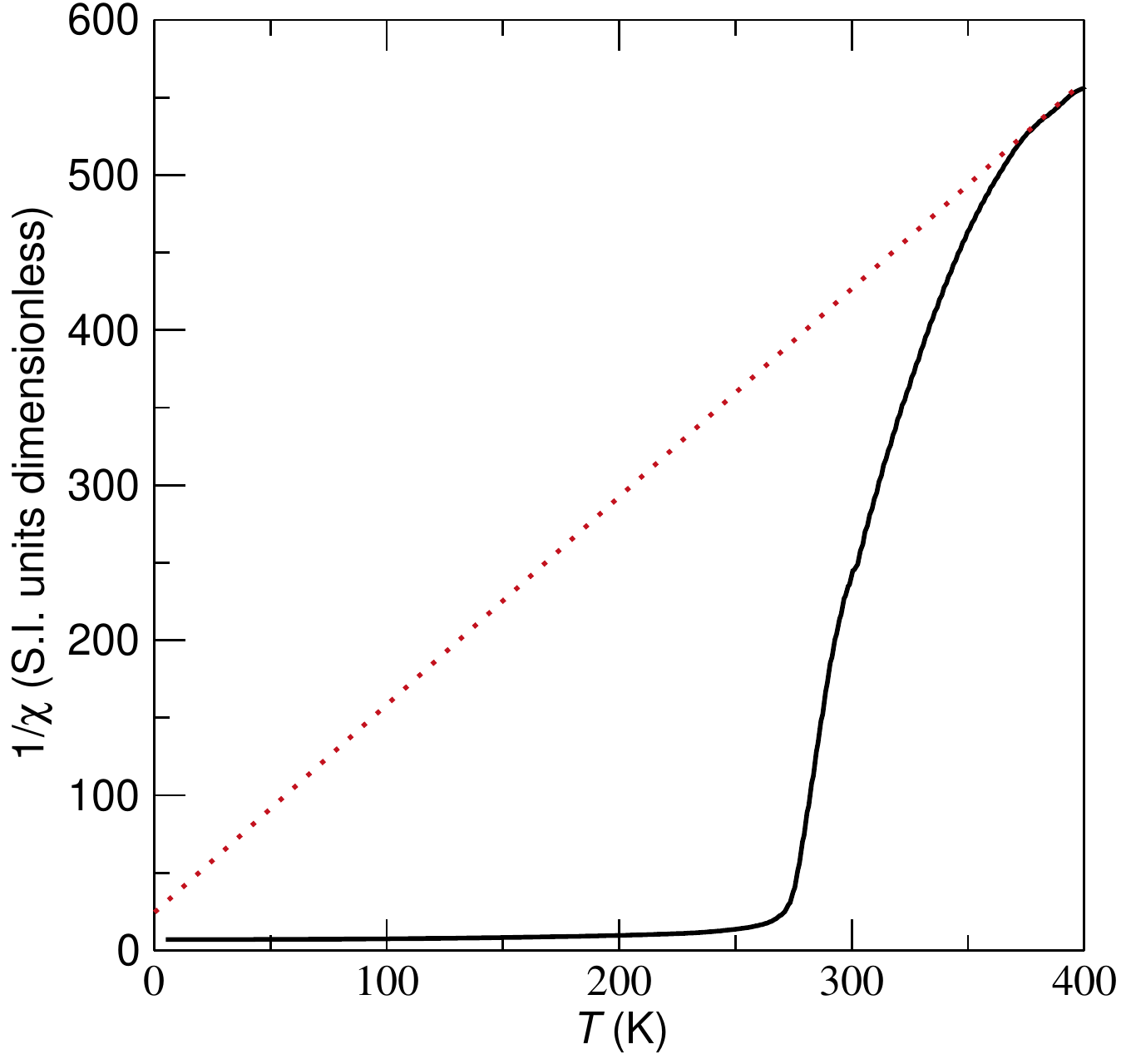} \\
\caption{\label{fig:inverse_susceptibility}
Inverse susceptibility of the field cooling curve in Mn$_3$As$_2$. There are not enough data points at the linear regime to provide a Curie-Weiss fit. The red dotted line indicates the extrapolation from the visible linear regime.
} 
\end{figure}

\FloatBarrier


\vspace{1em}
\begin{center}
\textbf{Cif file for Mn$_3$As$_2$ at 250~K}\\
\end{center}
\normalsize

\begin{verbatim}
data_As2_Mn3

# phase info for As2 Mn3 follows
_pd_phase_name  "As2 Mn3"
_cell_length_a  13.166866
_cell_length_b  3.681129
_cell_length_c  8.987757
_cell_angle_alpha  90
_cell_angle_beta   132.2013
_cell_angle_gamma  90
_cell_volume  322.708
_exptl_crystal_density_diffrn  3.3923
_symmetry_cell_setting  monoclinic
_parent_space_group.name_H-M_alt  "C 2/m"
_space_group_magn.name_BNS  "C 2/m"
_space_group.magn_point_group  2/m
loop_
    _space_group_symop_magn_operation.id
    _space_group_symop_magn_operation.xyz
     1  x,y,z,+1
     2  -x,y,-z,+1
     3  -x,-y,-z,+1
     4  x,-y,z,+1
     5  1/2+x,1/2+y,z,+1
     6  1/2-x,1/2+y,-z,+1
     7  1/2-x,1/2-y,-z,+1
     8  1/2+x,1/2-y,z,+1

# ATOMIC COORDINATES AND DISPLACEMENT PARAMETERS
loop_ 
   _atom_site_label
   _atom_site_type_symbol
   _atom_site_fract_x
   _atom_site_fract_y
   _atom_site_fract_z
   _atom_site_occupancy
   _atom_site_adp_type
   _atom_site_U_iso_or_equiv
   _atom_site_symmetry_multiplicity
Mn1    Mn2+ 0.30820     0.00000     0.68320     1.000      Uiso 0.011      4   
Mn2    Mn2+ 0.38883     0.00000     0.08690     1.000      Uiso 0.011      4   
Mn3    Mn2+ 0.00000     0.50000     0.50000     1.000      Uiso 0.016      2   
Mn4    Mn2+ 0.00000     0.00000     0.00000     1.000      Uiso 0.010      2   
As1    As   0.06069     0.00000     0.34317     1.000      Uiso 0.010      4   
As2    As   0.24682     0.00000     0.17768     1.000      Uiso 0.010      4   

loop_
   _atom_site_moment.label
   _atom_site_moment.crystalaxis_x
   _atom_site_moment.crystalaxis_y
   _atom_site_moment.crystalaxis_z
Mn1  0.0000      -0.52(9)    0.0000      
Mn2  0.0000      2.55(9)     0.0000      
Mn3  0.0000      -1.69(14)   0.0000      
Mn4  0.0000      0.59(6)     0.0000      

loop_  _atom_type_symbol _atom_type_number_in_cell
  As   8
  Mn   12

# Note that Z affects _cell_formula_sum and _weight
_cell_formula_units_Z  2
_chemical_formula_sum  "As4 Mn6"
_chemical_formula_weight  629.32

\end{verbatim}

\vspace{1em}
\begin{center}
\textbf{Cif file for Mn$_3$As$_2$ at 3~K}\\
\end{center}
\normalsize

\begin{verbatim}
data_As2_Mn3_mag_7

# phase info for As2 Mn3 follows
_pd_phase_name  "As2 Mn3"
_cell_length_a  13.085675
_cell_length_b  3.658847
_cell_length_c  17.832944
_cell_angle_alpha  90
_cell_angle_beta   132.1886
_cell_angle_gamma  90
_cell_volume  632.623
_exptl_crystal_density_diffrn  3.4609
_symmetry_cell_setting  monoclinic
_parent_space_group.name_H-M_alt  "C 2/c"
_space_group_magn.name_BNS  "C 2/c"
_space_group.magn_point_group  2/m
loop_
    _space_group_symop_magn_operation.id
    _space_group_symop_magn_operation.xyz
     1  x,y,z,+1
     2  -x,y,1/2-z,+1
     3  -x,-y,-z,+1
     4  x,-y,1/2+z,+1
     5  1/2+x,1/2+y,z,+1
     6  1/2-x,1/2+y,1/2-z,+1
     7  1/2-x,1/2-y,-z,+1
     8  1/2+x,1/2-y,1/2+z,+1

# ATOMIC COORDINATES AND DISPLACEMENT PARAMETERS
loop_ 
   _atom_site_label
   _atom_site_type_symbol
   _atom_site_fract_x
   _atom_site_fract_y
   _atom_site_fract_z
   _atom_site_occupancy
   _atom_site_adp_type
   _atom_site_U_iso_or_equiv
   _atom_site_symmetry_multiplicity
Mn1    Mn2+ 0.30820     0.00000     0.34160     1.000      Uiso 0.000      4   
Mn2    Mn2+ 0.38883     0.00000     0.04345     1.000      Uiso 0.000      4   
Mn3    Mn2+ 0.00000     0.50000     0.25000     1.000      Uiso 0.000      2   
Mn4    Mn2+ 0.00000     0.00000     0.00000     1.000      Uiso 0.000      2   
As1    As   0.06069     0.00000     0.171585     1.000      Uiso 0.000      4   
As2    As   0.24682     0.00000     0.08884     1.000      Uiso 0.000      4   

loop_
   _atom_site_moment.label
   _atom_site_moment.crystalaxis_x
   _atom_site_moment.crystalaxis_y
   _atom_site_moment.crystalaxis_z
Mn1  1.24(15)    -1.94(8)    2.27(11)    
Mn2  0.87(15)    2.28(7)     1.82(10)    
Mn3  0.0000      -4.48(13)   0.0000      
Mn4  -0.91(16)   1.21(6)     -0.56(11)   

loop_  _atom_type_symbol _atom_type_number_in_cell
  As   16
  Mn   24

# Note that Z affects _cell_formula_sum and _weight
_cell_formula_units_Z  4
_chemical_formula_sum  "As4 Mn6"
_chemical_formula_weight  629.32
\end{verbatim}